\documentclass[twocolumn,preprintnumbers,amsmath,amssymb,aps,pra,floatfix
]{revtex4}

\usepackage{isomath}

\usepackage{upgreek} 

\usepackage{epsfig, amsmath,amsfonts, amssymb,graphicx,amsthm,color}

\usepackage{mathtools}
\usepackage{dcolumn}
\usepackage{bm}
\usepackage{rotating}
\usepackage{amssymb}
\usepackage[latin1]{inputenc}
\usepackage{graphicx}

\usepackage{epstopdf}
\newcolumntype{M}[1]{>{\raggedright}m{#1}}

\begin{document}

\title{Real-time trajectory control on a deterministic ion source}
\author{C. Lopez}
\author{A. Trimeche}
\author{D. Comparat}
\author{Y.J. Picard}
\email[Corresponding author: ]{yan.picard@u-psud.fr}
\affiliation{Laboratoire Aim\'{e} Cotton, CNRS, Univ. Paris-Sud, ENS Paris Saclay, Universit\'e Paris-Saclay, B\^{a}t. 505, 91405 Orsay, France }

\date{\today}


\begin{abstract}
The major challenge to improve deterministic single ion sources is to control the position and momentum of each ion.
Based on the extra information given by the electron created in a photoionization process, the trajectory of the correlated ion can be corrected using a fast real time feedback system. In this paper, we report on a proof-of-principle experiment that demonstrates the performance of this feedback control with individual cesium ions.
The produced electron is detected with a time and position sensitive detector, whose information is used to quickly infer the position of the corresponding ion. Then the feedback system drives the ion trajectory through steering plates, the individual ion can thus be sent to any dedicated location. This enables us to perform deterministic patterning and reach a factor 1000 improvement in spot area.
The single ion feedback control is versatile and can be applied to different kinds of ion sources. It provides a powerful tool to optimize the ion beam and offers area for quantum systems and applications of materials science.
\end{abstract}

\maketitle


\vspace{-2mm}
\section{Introduction}
\vspace{-2mm}

The preparation and handling of individual particles
 is crucial for new technology development. 
Deterministic and high precision placement of individual atoms and ions offers 
  exciting prospects
for the realization of quantum based
 devices at the nanoscale \cite{bhushan2017springer,jamieson2017deterministic}.
 Several mechanisms for the
manipulation of individual atoms and ions have been demonstrated \cite{eigler1990positioning,Hor09,barth2010engineering}, but they are usually slow and non-deterministic.
Recently new type of ion sources have emerged: based either on cold trapped ions \cite{Sch09,schmidt2017nanoscopic}
 or cold trapped atoms  that are subsequently photoionized \cite{2016ApPRv...3a1302M}.
 From trapped ions, deterministic sources and
  implantation of several types of ions with down to 6 nm resolution have been demonstrated but the repetition rate and the beam energy range have still to be increased \cite{schmidt2016nanoscopic,schmidt2017nanoscopic,groot2019deterministic}.
From ionization of cold atoms, very high current and high brightness beam can be produced but with a non deterministic property \cite{2016Viteau,steele2017high}.
Progress toward
deterministic production of single ions has been made by exploiting the correlation between
an electron and associated ion after photoionization of cold trapped rubidium atoms \cite{sahin2017high} or an atomic beam \cite{mcculloch2018heralded}.
The electron
heralds the creation of a corresponding ion.
As stressed in ref. \cite{sahin2017high,mcculloch2018heralded}, the next step would be to extend the scheme beyond time-correlated feedback to have a position- and momentum-correlated feedback so as to provide total control over each single ion in a beam. This would provide
general and powerful tools to optimize the ion beam and 
would offer a complete  portfolio  for
quantum optics and applications in materials science.

\begin{figure}[h!]
	\centering
	\includegraphics[width=1.0\linewidth]{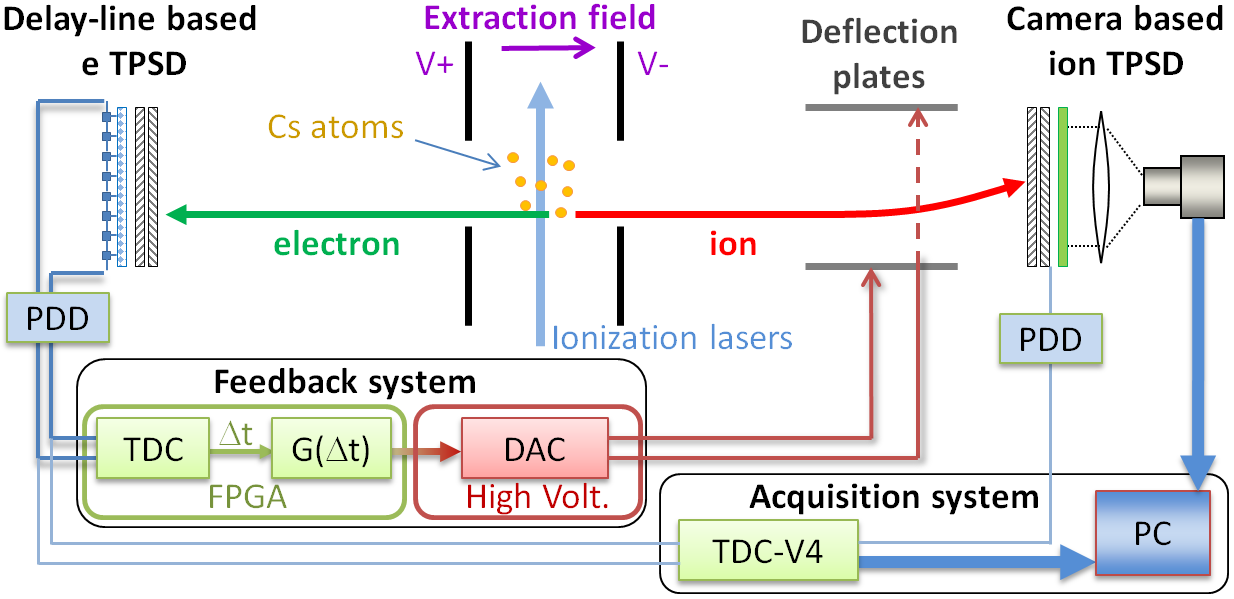}
	\caption{Scheme of the experimental setup. Electrons and ions, produced by photoionization of a cesium beam, are extracted in opposite directions and detected by two time and position sensitive detectors (TPSD). The acquisition system handles the information from the detectors and processes the data. The feedback system controls the ion trajectory by acting on the deflection plates with the appropriate voltages for each ion (presented here in 1D, see text for more details).}
	\label{fig:experimental-apparatus}
\end{figure}

This  position correlated feedback is exactly the step that we demonstrate in this paper. We exploit the correlation between
an electron and the associated ion created by photoionization of an atom. We use fast real-time feedback to modify the ion trajectory  based on information obtained by direct detection of the electron.
We demonstrate an active correction of the trajectory allowing the ion to be sent to any desired final location independently of its initial position.  This allows a full control of the final spot pattern with increased accuracy compared with the uncorrected one, which enables us to reduce the ultimate achievable spot size by correcting each ion trajectory.
A three orders of magnitude reduction in spot area is demonstrated in a one-by-one ion driving experiment on a millimeter initial size ion spot. In principle this reduction remains valid even if we start from a micrometer spot size. In this case the nanometer scale is probably reachable using for instance a focused ion beam (FIB) column. 
This trajectory correction can be generalized to control all beam parameters (position, momentum, energy) and thus can greatly improve the phase space density, the brightness and the emittance of the ion beam. In this paper we focus only on basic correction schemes consisting of the correction of the ion position on the target detector and the application of virtual diaphragm and mask.
In the following, we first present the apparatus. We then study the correlation between electrons and ions. Finally, we present the results of the feedback correction allowing us to create any patterns of ion spots.

\vspace{-2mm}
\section{Experimental setup}
\vspace{-2mm}
\subsection{Ionization and extraction}
\vspace{-3mm}

The setup is based on a double time-of-flight (TOF) spectrometer with time and position sensitive detectors (TPSD) at opposite ends monitored in coincidence mode (see Fig. \ref{fig:experimental-apparatus}).
For our experiment, we use cesium (Cs) atoms so as to create ion-electron pairs. Cs effuses from an oven and passes through a 3~mm-diameter and 10~cm-length heated copper tube, and then propagates 20~cm to the ionization region. Photoionization is performed in a quasi uniform static electric field produced in between two holed electrodes (4~mm hole diameter) separated by 10~mm. 
By use of narrowband lasers for the photoionization, Doppler selection can be performed to reduce drastically the effect of the effusive atomic beam velocity dispersion.
Ionization of Cs atoms is performed with a three-photon transition process where
the first transition uses a horizontal laser beam counter-propagating with respect to the atomic beam. It is
ensured by a 852~nm laser beam that excites atoms from the $6S_{1/2}$ F = 4 level to the $6P_{3/2}$ F = 5 level. For Doppler compensation, this laser is locked by saturated absorption on a vapor cell and detuned by an acoustic optical modulator. 
The second transition is performed by a perpendicular 1470~nm laser beam coupling the excited $6P_{3/2}$ state to the $7S_{1/2}$ state. This laser is locked thanks to a Cs cell  excited by the 852~nm  laser. The last transition is ensured by a tunable Ti:sapphire  laser, in the same plane but tilted 45${^\circ}$ from the previous lasers.  It excites the atoms from the $7S_{1/2}$ state to a Rydberg state or the ionization continuum depending on the chosen wavelength (770-795~nm). 
This geometry enables us to control the ionization volume at the crossing of the laser beams.
These lasers are focused, inside the extraction zone, with a typical size of $\simeq 500\,\mu m$. 
For the results presented here, ionization occurs in a 2200~V/cm field and the (vacuum) Ti:sapphire laser wavelength is tuned to 794.432~nm.
Fine adjustments are made to the Ti:sapphire laser wavelength to optimize the detected electron and ion signals. We choose this wavelength because it results in an efficient excitation towards a Rydberg state that autoionizes, and produces an efficient source of ion and electron pairs.
Upon ionization, electron and ion are accelerated by the 2200~V/cm static electric field in opposite directions toward the TPSD located 355~mm (for electrons) and 285~mm (for ions) from the ionization region. Typical TOF values are about 20~ns for the electrons and 7~$\mu$s for the ions. Outside the ionization region, along the ion trajectory, a set of horizontal and vertical deflection plates are dedicated to the ion trajectory correction. 

\vspace{-3mm}
\subsection{Detection and acquisition}
\vspace{-3mm}

To reveal the correlation between electron and ion and use it to infer the ion position, we need to determine the position and TOF of both particles. For this, we use two TPSD monitored in coincidence mode by an acquisition system that works independently of the fast feedback system (see Fig. \ref{fig:experimental-apparatus}).
The electron TPSD is composed by a set of imaging quality microchannel plates (MCP) and a delay line detector (DLD). The DLD is based on capacitive coupling between a resistive anode, that collects the electron cloud arising from the MCP, and an encoding surface through a ceramic plate \cite{ceolin2005high}. This detector has a position resolution of about 50~$\mu$m. The two pairs of analog signals coming from X (horizontal) and Y (vertical) delay lines pass through homemade fast amplifiers and peak-detection discriminators (PDD \footnote{DTPI, ISMO / LUMAT, CNRS / Univ.Paris-Sud / IOGS / Univ.Paris-Saclay, Orsay, France.}) that deliver short ($\sim$15~ns) logical signals. These signals end in a multi-channel time-to-digital converter (TDC-V4 \footnote{DTPI:  https://www.pluginlabs-universiteparissaclay.fr/fr/entity/917038-dtpi-detection-temps-position-image}) with a resolution of 60~ps (rms). 
The ion TPSD is composed of a set of imaging quality microchannel plates (MCP), a phosphor screen and a CMOS camera. It is based on correlation between the brightness of the spot on the phosphor screen and the amplitude of the time signal on the MCP \cite{urbain2015zero}. This detector has a position resolution of about 50~$\mu$m. 
The two time signals for ion and electron generated by the MCP are digitized by PDD and recorded by the TDC-V4. This enables us to measure the relative TOF of the detected ion and electron. The time and position signals delivered by the two TPSD are handled by a C++ monitoring home-written program that controls acquisition and stores the data. For each detected event, the acquisition program determines the position and relative TOF of the ion and the electron. If this TOF fits in a given range ($\simeq 200$~ns corresponding to the spread caused by the potential difference across the ionization region) around the expected value, the pair is considered as coming from a unique ionization event and is labeled coincident in time. This is a mandatory condition to obtain a electron/ion correlation. For this proof of principle experiment, a low counting rate ($\simeq 1$~kHz) has been chosen in order to be compatible with the acquisition rate of the CMOS camera. This rate ensures that the false coincidences are negligible and that the correction by the steering plates never concerns more than one ion.

\begin{figure}
	\centering
	\includegraphics[width=1.0\linewidth]{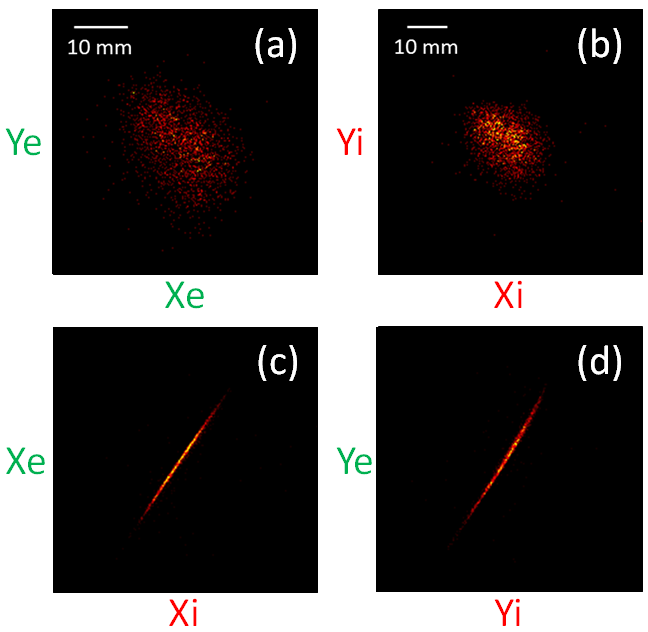}
	\caption{(a) and (b): raw images from the position sensitive detectors of electrons (Xe,Ye) and ions (Xi,Yi) respectively; (c) and (d): correlation patterns between
respectively the horizontal and vertical coordinates of the coincident events. The position scales are shown on (a) for (Xe,Ye) and (b) for (Xi,Yi) and are the same for all the images presented in this paper.}
	\label{fig:conincidence}
\end{figure}

The acquisition system is used to characterize the correlation between the two particles produced by the ionization.
Figure \ref{fig:conincidence} shows typical correlation between individual ion and electron arising from all the validated coincident events.
The images of all electrons and ions detected in coincidence are shown in Figure \ref{fig:conincidence} (a) and (b) respectively for several thousand acquisition events. For each coincident event, the X (resp. Y) coordinate of the electron is plotted with respect to the X (resp. Y) coordinate of the corresponding ion on the correlation map seen in Figure \ref{fig:conincidence} (c) (resp. d). These maps show two narrow correlation patterns on each coordinate. The slight curvature of these correlation patterns is due to aberrations such as the one induced by the residual magnetic fields seen by the electrons during their flight. Hence, for each detected electron, and from the measurement of its position, we can deduce the position of the coincident ion. 
The correlation pattern must be one-to-one (injective) and its width limits the precision of the inferred position of the ion. This width is mainly 
affected by non-zero electron emission energy, ion velocity spread, extracting field inhomogeneity and a variety of other minor factors such as fluctuation of the fields affecting the trajectories, resolution of the electron detector. The main factors are overcome respectively by threshold ionization (electron emission energy on the order of few meV), Doppler selection (ion velocity spread of the order of few m/s) and spatially resolved Rybdberg state excitation inside the ionization zone (giving a spatial ionization width of  $10-100~\mu$m).

\vspace{-3mm}
\subsection{Fast feedback system}
\vspace{-3mm}

The fast determination of the detected electron position allows us to perform an active feedback control in real time for each ion trajectory.
The feedback system gets X and Y coordinates of the electron from the DLD and processes them to apply the correction by sending appropriate voltages onto the four deflection plates. In more detail, copies of the logical signals arising from the DLD are sent to a field-programmable gate array (FPGA) \cite{FPGA1,FPGA2} that manages the feedback calculation experimentally optimized. Inside this integrated circuit, four TDC measure the arrival time of each signal. The TDC are of tapped delay line type \cite{favi200917ps} with a standard deviation of $\simeq$~75~ps.
Thus, a Finite State Machine within the FPGA checks the validity of each event. If it gets the four expected signals from the DLD  during a window corresponding to the delay-line propagation time ($\simeq$~80~ns), the event is validated. If not, multiple hits or incomplete hits are ignored.
Once the event is validated, the X and Y coordinates of the electron are extracted from the time difference $\Delta t$ of each pair of signals and are processed by two transfer functions $G(\Delta t)$ (see Fig. \ref{fig:experimental-apparatus}) to deliver the correction we want to apply.
After this FPGA calculation, a high-speed isolated serializer transfers numerical data to a high-voltage electronic board with a delay of 400 ns. On this board, four digital to analog converters (DAC) with amplifiers generate the appropriate voltages (offset voltage pulses of few tens of volts during few $\mu$s) to supply the steering plates, with a transition time on the order of $\mu$s. The correlated ion reaches the deflection plates after a TOF of $\simeq 3~\mu$s which is longer than this voltage pulses transition time.
When the ion is in between the steering plates, the applied correction voltages deflect its trajectory according to the feedback system instructions.
Additionally, to enhance the deterministic behavior of our source, we set by default the deflectors in a configuration where the ions are permanently steered out from the target, except when an electron is validated by the feedback system which then lets the correlated ion pass through with the adequate correction. This technique makes our ion source heralded.

\begin{figure}
	\centering
	\includegraphics[width=1.0\linewidth]{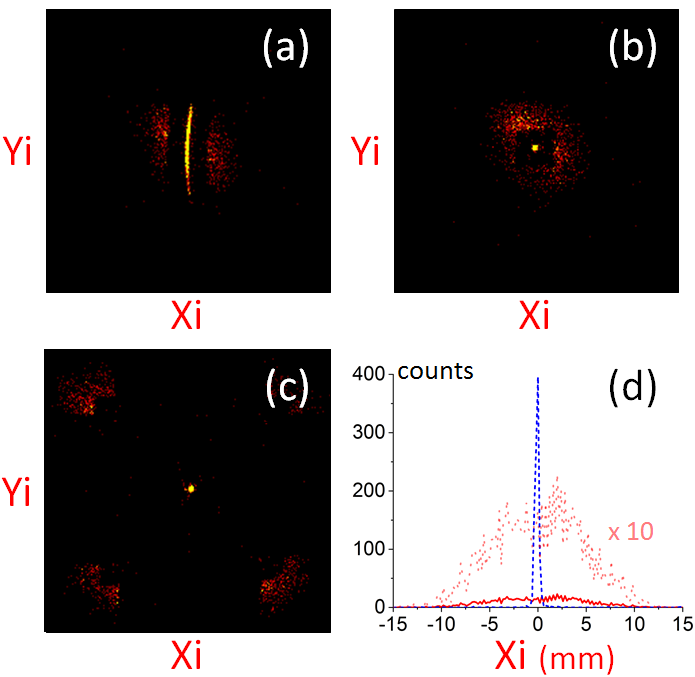}
	\caption{Raw images from the ion detector after: (a) 1D correction along X coordinate on a selected zone; (b): 2D correction on a selected zone; (c): 2D correction of selected ions with non-selected ions pushed away from the target; (d) projection on the X-axis of the ion position distribution without correction (solid red and dotted light red for amplitude $\times 10$) and with the 2D correction (dashed blue). The FWHM is reduced by a factor of $\sim~30$ on each direction.}
	\label{fig:corrections1D2D}
\end{figure}

\vspace{-2mm}
\section{Results and discussion}
\vspace{-2mm}

Figure \ref{fig:corrections1D2D} shows the key control configurations we apply to verify the validity of our approach and to demonstrate the performances of our feedback system. Figure \ref{fig:corrections1D2D}  (a) presents an image of the detected ions after applying a 1D-correction on the X-axis. As can be seen, the corrected ions are aligned along the Y-direction.
Here a simple linear transfer function $G(\Delta t)$ is set along the X component so that each ion is carried onto a vertical line when it reaches the target detector. The slight curvature seen in the resulting structure is due to the non-linearity of the correlation patterns. As mentioned before, this non-linearity is mainly caused by the residual magnetic field. The feedback system could correct this curvature but for this paper we ignore it and show a simple way to bypass it by applying a virtual diaphragm that selects the quasi-linear part.
Figure \ref{fig:corrections1D2D}  (b) shows an image of the detected ions after application of a 2D-correction on the X and Y axis. As can be seen, the corrected ions are sent to a very small spot on the target detector. Note that
this spot is much smaller than what we obtain at best by focusing with our electrostatic lenses thanks to the one by one ion trajectory correction. 
A non linear correction is very possible with this feedback system. This will lead to an even smaller spot but it is beyond the scope of this paper.
Note that in these experiments we select a 1D zone (Fig. \ref{fig:corrections1D2D} (a)) or a 2D zone (Fig. \ref{fig:corrections1D2D} (b)) in which we apply the correction.
If the validated electron coordinate X or Y is inside the zone of interest, the steering voltages are set to the appropriate values to correct the ion trajectory. If outside, these voltages are set to zero so that the ion passes through without deviation. An interesting alternative is to push out this non-selected ion.
Figure \ref{fig:corrections1D2D} (c) shows an experimental result for this configuration, where we can see at the center of the image the corrected ions spot, and at the edges of this image the pushed ones. Here we do not push out the non-selected ions very far from the zone of interest so as to see them, but we can easily drive them far away from the detector target. This corresponds to add a virtual diaphragm to the correction process.
Figure \ref{fig:corrections1D2D} (d) shows the profile of the ion spots with and without correction. It demonstrates a reduction by a factor of 30 of the spot width. The same result is obtained on the other axis, which gives a three order of magnitude reduction in the ion spot area after correction. By numerically simulating the correction, using the raw data from the acquisition system without correction, we obtain the same result which means that this reduction factor is more limited by the width of the correlation pattern than by the feedback system accuracy. The virtual diaphragm functionality is thus very useful. For instance when the width of the correlation pattern is not uniform, we can select a zone with the thinnest width in which we apply the correction and remove the non-selected ions. This leads to an improvement of the correction quality without a large current loss. This virtual mask enables us also to select a zone which correspond to a very small part of the ionization region in order to minimise the effect of the energy spread and aberrations.

\begin{figure}
	\centering
	\includegraphics[width=1.0\linewidth]{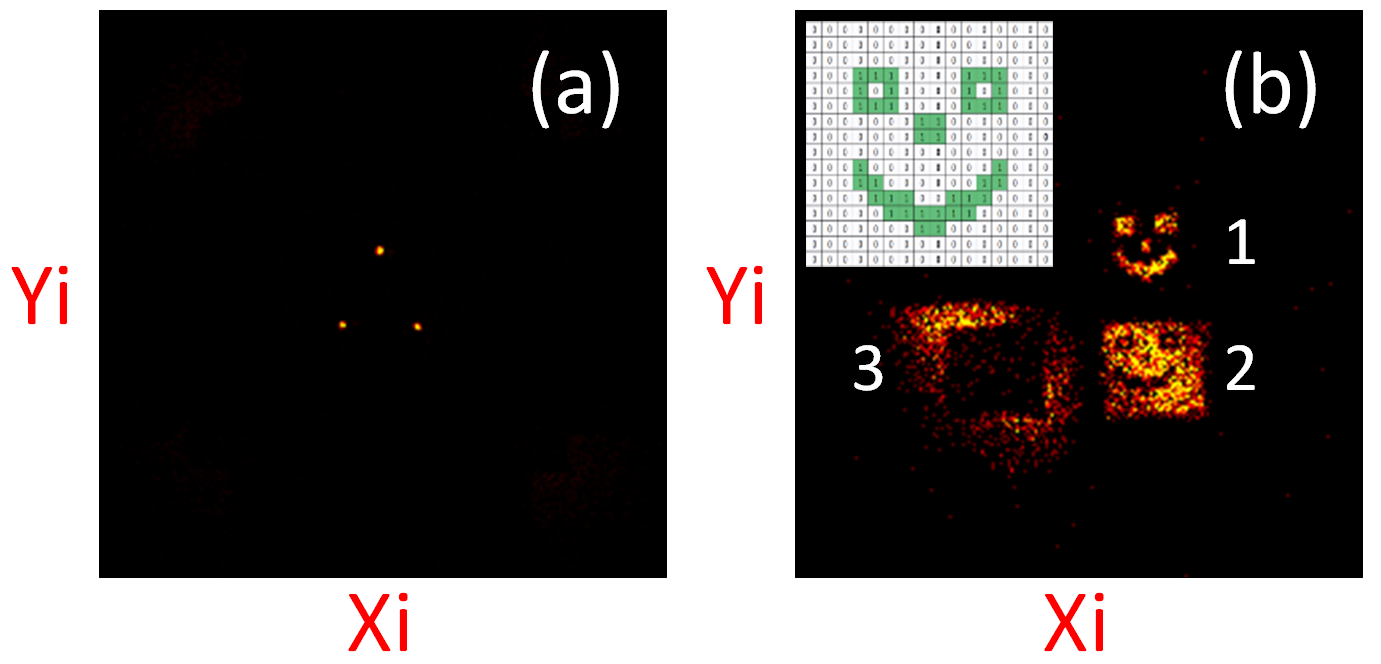}
	\caption{Raw images from the ion detector after: (a) application of a 2D correction in which the ions are carried on to three different spots at the target; (b): application of a virtual mask implemented into the FPGA.}
	\label{fig:patterncorrections}
\end{figure}

Our feedback system is versatile and allows us to realize more complex processes. Figure \ref{fig:patterncorrections} (a) shows an example of a complex pattern that can be produced, where the selected ions are carried on to three different spots as a function of the position of their correlated electrons. Here the selected zone is separated into a $2\times 3$ array, but more complex patterns can easily be done by using larger arrays. Indeed, we can draw any desired form on the target with the corrected ions. As we have seen in figure \ref{fig:corrections1D2D} (c), the feedback system can play the role of a diaphragm. With a more complex selected zone, and without correction on the selected ions, the feedback system can also play the role of a mask.
An example of such a mask is shown in figure \ref{fig:patterncorrections} (b) where: a $16\times 16$ array mask implemented into the FPGA is represented in the upper-left corner; the selected ions corresponding to the green (resp. white) pixels are placed in (1) (resp. (2)) on the detector target; the non-selected ions are pushed away to (3). Note that the structures (1,2,3) are accumulated during the same experiment and are all kept inside the detector target for demonstration purposes.
Obviously, if we couple a correction of the selected ions with this kind of virtual mask, we can drive single ions to any desired location and realize any conceivable pattern on the target. 

\vspace{-2mm}
\section{Conclusion}
\vspace{-2mm}

In conclusion, by using the correlation between ion and electron, produced by photoionization of an atomic beam, we develop 
a fast feedback system to control the ion trajectory.
Compared with previous Rb photoionization experiment \cite{sahin2017high,mcculloch2018heralded} we add, to the deterministic character of the ion delivery, the control of its trajectory and complex patterning such as multi-spots and virtual diaphragm/mask.
The focusing performance of our system is better than those obtained using conventional lenses due the one-by-one ion trajectory correction.
This correction method is quite general and can be implemented in different setups and the feedback can be applied not only to steering plates but also to any electromagnetic components that affect the ion trajectory (grids, lenses, prisms, $etc$...).
This feedback system can even correct more complex effects such as finite source size, electrostatic or magnetic lenses aberrations, stray fields, $etc$... by applying non-linear and/or individually adapted corrections for each ion.
In principle such a system can be used in reverse mode to control the electron instead of the ion. This only requires a dedicated (acceleration-deceleration) geometry that ensures the detection of the ion before the electron.
This proof-of-principle experiment is performed with a $mm$ scale ion source that is downsized to $\mu$m scale on the target at $\simeq$~1~kHz repetition rate. 
Coupled with appropriate setup, this feedback system may produce 10 MHz rate (1 pA) beam by reducing the ion TOF inside the steering plates and increasing the capacity of the detector to handle high count rates.
Numerous methods such as discreet anodes detectors, coherent excitation process and Rydberg dipole blockade can probably be used to increase the rate even further \cite{PhysRevA.84.023413,PhysRevA.94.023404}.
Spectacular improvement of the final spot resolution can be also achieved using a FIB column based on direct photoionization or excitation of field ionized Rydberg states of ultra-cold atoms \cite{2016Viteau,2016ApPRv...3a1302M}. This allows a drastic reduction of the initial ion velocities leading to a very narrow ion/electron correlation pattern. Preliminary simulations prove that our feedback system can produce sub-nm spots even for quite low beam energy since we correct each ion independently.  The deterministic behavior and  complex patterning obtained by this feedback method makes it very useful for FIB studies such as for nano-structuring, microscopy and surface spectroscopy \cite{Orloff2008,Gie09}.
 This can be used in future semiconductor fabrication or in quantum technology development for instance by structuring  devices at an unprecedented precision  \cite{jamieson2017deterministic,tosi2017silicon}.
Hence, this feedback method offers  opportunity for ion beam
techniques such as aberration correction, high resolution imaging or accurate implantation.

\section{Acknowledgements}
\vspace{-3mm}

This work was supported by the Fond Unique Interminist\'eriel (IAPP-FUI-22) COLDFIB, the
European Research Council under the grant agreement No 712718-LASFIB, ANR HREELM and CEFIPRA No. 5404-1.
The authors thank G. Chaplier and DTPI for useful help in developing the detectors and L. Calmon for her contribution to this work.

\bibliographystyle{h-physrev}
\bibliography{biblio,biblio_add}

\end{document}